# Early quantum computing applications on the path towards precision medicine

*Frederik F. Flöther[1]*
[1]*QuantumBasel, Schorenweg 44b, 4144 Arlesheim, Switzerland*

**A quantum state of mind**

Quantum computing is one of the most recent arrivals in medicine's toolbox although quantum theory and medicine have arguably been entangled ever since Schrödinger's cat[1]. Quantum computing may turn out to be one of the toolbox's most powerful instruments; it follows in the footsteps of the information processing revolution, which transformed our world and medicine with it. This includes computational breakthroughs over the last decades such as medical imaging, quantitative structure-activity relationship (QSAR) models, computer-aided drug design, and genomic sequencing.

While many new technologies are labelled "exponential" nowadays, quantum computing might be the only one fully deserving of that label. It is the only known computational model that can achieve exponential speedups over classical computers[2]. Quantum computing is based on fundamentally different hardware and software. Instead of computing with bits, in the gate model quantum bits (qubits) are manipulated with quantum gates, thus forming quantum circuits and algorithms. Such qubits can be realized in different physical systems, including atomic nuclei, atoms and ions, electron spins, photons, semiconductor defects, and superconducting circuits[3].

Quantum algorithms are designed to take advantage of quantum mechanical laws and properties – such as quantum entanglement, interference, and superposition – to provide speedups over their classical counterparts. These properties are generally counterintuitive, their meaning and interpretation still hotly debated and providing deep inspiration in metaphysics and philosophy 100 years on from the original development of quantum theory[4]. From a technological point of view, with the addition of each qubit there is a doubling of the quantum state space, which mathematically represents the qubits' states; as quantum computers can manipulate all states at once, this results in a doubling of the potential computational power. Hence, the technology is inherently exponential.

Nevertheless, while one may be rightly excited by the arrival of such an extraordinary tool, a range of subtleties and caveats must be appreciated. Although the field of quantum algorithms is still of a young age and continues to be intensely researched, not every classical algorithm and application is expected to benefit from a speedup via a quantum counterpart[5]; many computational tasks, certainly the arguably more mundane ones involved in, say, adding two numbers or sending an email, likely lack quantum speedups. This is why quantum computers are not expected to replace classical ones but, instead, work hand-in-hand with them in hybrid computational pipelines. Furthermore, there are many intricacies involved in using quantum systems. For example, given the fragility of quantum states, sophisticated error suppression, mitigation, and eventually correction is essential[6]. In addition, reading in, encoding, and reading out data, particularly classical data, is not trivial for a quantum computer. Large volumes of classical data cannot (yet) be efficiently read in/out ("input/output problem") by quantum computers. Techniques such as dimensionality reduction can alleviate these types of issues but thinking of today's quantum computers simply as big data machines would be a mistake.

The field has advanced so rapidly over the last years that the consensus regarding the impact of quantum computers has now shifted from "if" to "when/how"[7]. A wide range of early use cases and applications are already being explored with quantum computers across industries through proof-of-concept studies (Figure 1), including in healthcare, life sciences, and medicine. For medical researchers and practitioners, there is ample reason to be excited by the quantum era and one can already begin taking the first steps today towards getting familiar with the technology and into a quantum state of mind.

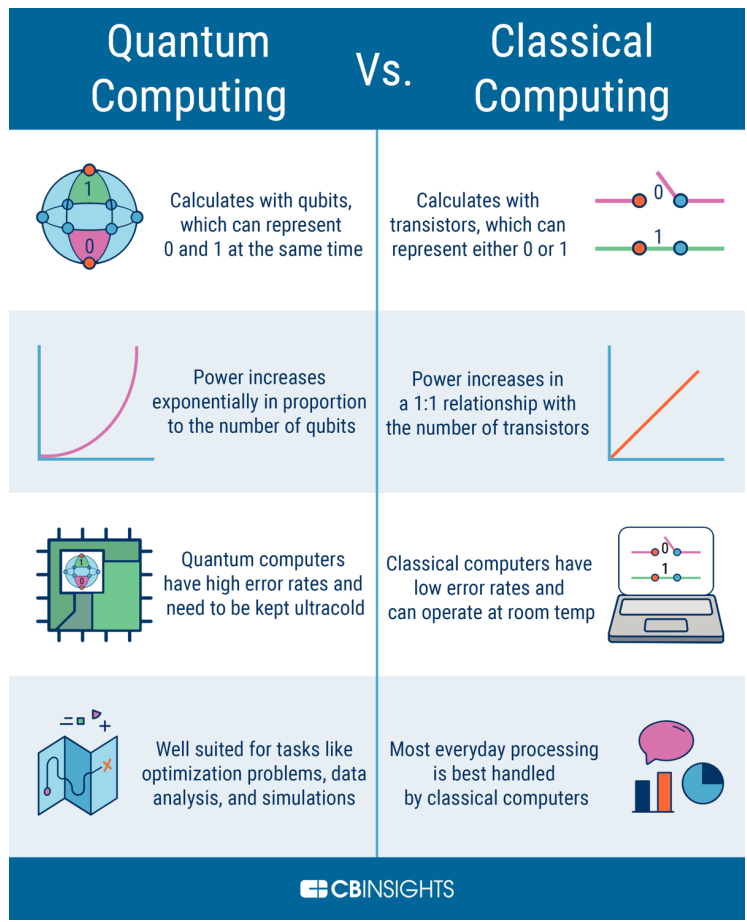

*Figure 1 (source[8]): Quantum computing is fundamentally different from classical computing. It may end up being one of the most powerful instruments in medicine's toolbox, enabling key advances in our quest to unlock the complex secrets of biology.*

**Applying quantum computing in medicine**

When exploring potential quantum computing use cases, several criteria must be considered (Figure 2); the use cases discussed in this chapter were selected based on such criteria and on the availability of published proof-of-concept studies. First, as mentioned before, quantum computers are not silver bullets that can solve every problem faster. It is thus imperative to consider the quantum feasibility of a given use case, that is, which quantum algorithms are applicable and what speedup can be expected. Some of these methods may be suited to the noisy intermediate-scale quantum (NISQ) systems that we have today (often with limited or no speedup guarantees), others may only be applicable in the future era of fault-tolerant quantum machines (often with speedup guarantees, in some cases even exponential ones). Second, the impact of a given use case must be evaluated, taking into account the business and scientific value it could generate. This may not be straightforward, but rough estimates such as "increasing the medical image classification accuracy by X % would engender diagnoses to be made Y % earlier, with associated improvements in patient outcomes and reductions in treatment costs" can already give useful indications. Third, applicable prerequisites must be studied; the ease with which they can be fulfilled often varies strongly by organization. These could include, for instance, availability of specific data with a minimum quality, existence of specialized skills in the team, and access to certain quantum and classical systems.

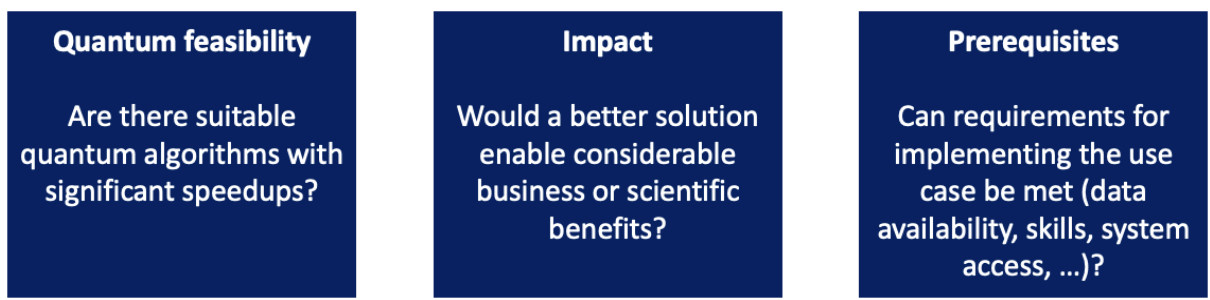

*Figure 2: Multiple criteria must be considered when evaluating quantum computing use cases.*

Defined by the characteristics of the algorithms and the types of problems for which the algorithms are typically employed, three primary quantum algorithm application categories can generally be distinguished:

1. Simulating nature – including chemistry, material science, physics
2. Processing data with complex structure – including artificial intelligence / machine learning (AI/ML), factoring, ranking
3. Search and optimization – including pricing, risk analysis, sampling

Each of these categories has multifarious use cases across industries, including in health and medicine where dozens of proof-concept experiments have already been conducted[9]. Depending on the use case, the quantum benefit may be speed or accuracy or handling noisy input data sets or solving a problem that had been hitherto intractable.

Precisely when applications based on algorithms in each category become commercially viable is a subject of much debate. One view is that algorithms that simulate nature will bring the first benefits due to their natural fit with quantum mechanics; Richard Feynman, one of the pioneers of quantum theory, famously put it in 1981 as "…if you want to make a simulation of nature, you'd better make it quantum mechanical…"[10]. Another view is that quantum machine learning and optimization will lead to the first breakthroughs due to the broad range of techniques and applications in these fields and the global cross-industry interest in data science methods.

The healthcare and life sciences sectors have seen rapid growth in quantum studies over the last years; drug discovery has been one focus area[11]. An increasing number of these are even run on real quantum hardware. Major investments have been made; for example, the Novo Nordisk Foundation Quantum Computing Programme, launched in collaboration with the University of Copenhagen, has over $200M in funding over 12 years[12]. Multiple germane reviews exist, focusing on the computational biology perspective[13,14,15,16,17,18]. In addition to pharmaceutical and life sciences uses[19], clinical applications and medical use cases are now coming increasingly into the picture[20]. The present chapter focuses on three key use case areas associated with (precision) medicine (Figure 3).

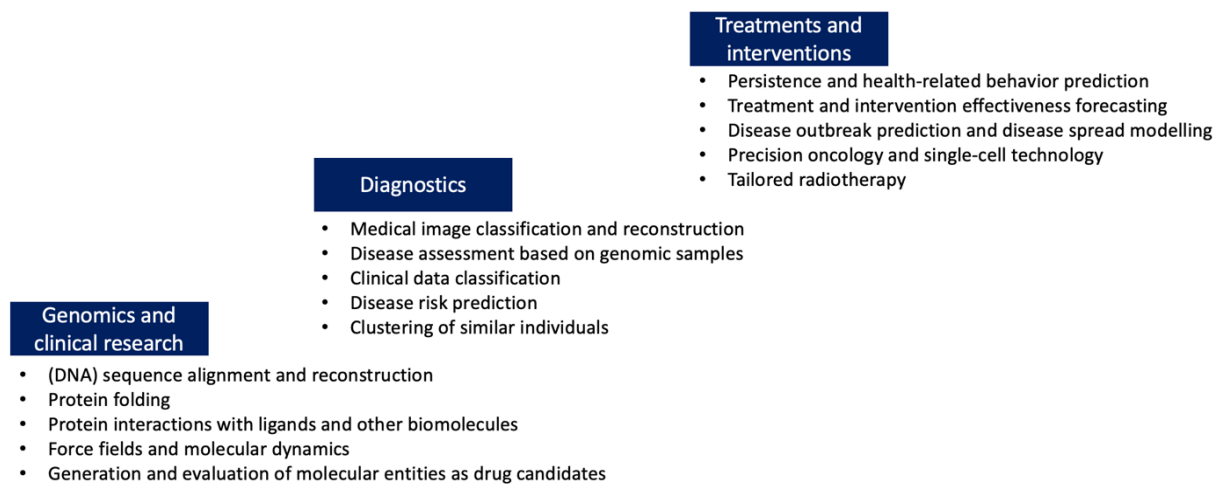

Figure *3 (adapted from source[9]): Examples of quantum computing use cases in medicine grouped into three areas.*

For some applications, quantum computing's expected benefits will drive precision medicine in a stricter sense, making it more granular and proactive; for other applications, quantum computing is likely to accelerate precision medicine more indirectly, for instance, by helping discover new drugs or therapies that benefit large groups of individuals. Note that there are also quantum computing use cases in the operations associated with medicine, including the optimization of supply chains; these are beyond the scope of this chapter.

Furthermore, the focus of this chapter is on universal quantum computing, which allows all quantum (and classical) algorithms to be implemented, simulated, or approximated[21]. Alternative computing approaches exist, including quantum annealing (a technique for solving optimization problems that leverages quantum fluctuations[22]) and quantum-inspired methods (classical algorithms that were developed using quantum mechanical ideas). Quantum annealing, for example, has also already been applied to a range of optimization problems in health and medicine across the areas shown in Figure 3. These include proof-of-concept experiments in genome assembly[23,24], binding affinities between proteins and DNA targets[25], peptide design[26,27], molecular similarity[28], intensity-modulated radiation therapy[29], and nurse scheduling[30].

Likewise, other quantum technologies are already being explored for medical applications but are not covered in this chapter[31]. An example is quantum sensing, where quantum properties are leveraged to achieve enhanced sensing precision[32]. Another example is quantum security, where new security threats and opportunities are emerging[33]. These respectively include, for instance, the possibility for adversaries to steal encrypted data today in order to decrypt them years from now with powerful quantum computers and the emergence of quantum-safe encryption techniques (which may be fully classical or may leverage quantum properties, as is the case in quantum key distribution).

Each of the three main use case areas will now be discussed in turn, including examples of proof-of-concept studies that have already been conducted by different organizations.

*Genomics and clinical research*

Approximately 40 % of diseases have a genetic component[34]. As such, genomics forms the foundation for precision medicine. In genomics and clinical research, all three of the aforementioned quantum algorithm application categories play prominent roles. This is because it is essential to understand the molecular behavior and the chemistry of small and large biomolecules as well as analyze, search, and optimize the complex patterns found in omics and phenotype data.

For example, many omics applications require sequence alignment, where sequences are compared in order to identify degrees of similarity. Such problems quickly become exponentially complex[35]. Researchers from the Aristotle University of Thessaloniki showed that quantum algorithms can potentially run faster, while requiring fewer computational resources, for pairwise sequence alignment[36]. A related problem concerns the (de novo) reconstruction of DNA after it was broken into fragments and sequenced; quantum computing was explored in this space by researchers from the Delft University of Technology and the University of Porto[37].

Proteins and their interactions represent another field of high interest. Classical AI/ML methods have made tremendous progress in being able to predict protein structures from amino acid sequences[38,39]. Nevertheless, many problems remain intractable, for instance, forecasting the structures of proteins with unnatural amino acids as well as understanding the interactions of proteins with ligands, water, and other biomolecules. Quantum techniques have shown early promise in eventually enhancing protein structure and interaction predictions. Researchers from ProteinQure[40], as well as IBM and Sorbonne University[41], demonstrated how this may be done for lattice-modelled proteins; a collaboration between Tencent and Roche[42] considered extensions to non-lattice-modelled proteins. Protein-ligand interactions were studied with quantum methods by QC Ware and Boehringer Ingelheim[43] as well as Quantinuum and Roche[44]. The quantum (and classical) resources required to assess the electronic structures of cytochrome P450 enzyme active sites were estimated by an international collaboration consisting of researchers from Google, Quantum Simulation Technologies, Columbia University, Boehringer Ingelheim, and the University of Innsbruck[45].

Moreover, in the context of accelerating drug discovery, the ability to run efficient molecular dynamics calculations is key. These, in turn, require accurate knowledge of molecular force fields. Quantum computing may improve these types of computations, as was explored by a collaboration between IBM, ETH, and CERN[46]. Calculations of energy profiles as well as bond interactions were carried out through an integrated quantum-classical pipeline by Tencent, AceMapAI, and the Ningbo University of Technology[47]. In early drug discovery, one must also downselect the thousands or millions of virtual drug candidates to be left with those which have activity towards a protein target while exhibiting favorable absorption, distribution, metabolism, excretion, and toxicity (ADMET) properties. Quantum algorithms were shown to be competitive with classical approaches for such virtual screening in a study by the Hartree Centre and IBM[48]. Similarly promising results were found in a toxicity screening experiment by researchers from PASQAL, Sorbonne University, and Paris-Saclay University[49]. Even generating new molecular entity candidates through quantum AI/ML has been investigated by Pennsylvania State University and IBM[50].

*Diagnostics*

Early and accurate diagnoses are a cornerstone for optimal outcomes. For instance, survival rates increase by a factor of 9 and treatment costs decrease by a factor of 4 when colon cancer is diagnosed early[51]. From the algorithm application categories, quantum AI/ML methods are particularly relevant to diagnostics due to the abundance of problems that require complex patterns and correlations to be analyzed and predictions to be made.

Medical images represent a key modality for such assessments; their interpretation has become increasingly important for correct diagnoses. This includes magnetic resonance imaging (MRI) and X-ray pictures. For instance, QC Ware, the Institut de recherche en informatique fondamentale, the Indian Institute of Technology Roorkee, and Roche classified chest X-ray and retinal color fundus images using quantum neural network algorithms, exploring the potential for speed and accuracy improvements compared with purely classical methods[52]. As another example, researchers from the University of Porto also used quantum neural networks to classify full-image mammograms as malignant or benign[53].

In addition to medical images, quantum AI/ML has been applied to a variety of other clinical and real-world data sets for the purpose of diagnosing and predicting disease risks. For instance, the University of Virginia classified Alzheimer's disease by applying quantum techniques for comparing single-cell neuronal genomic copy number variations[54]. In another initiative, the Medical University of Vienna, Johannes Kepler University Linz, and the Software Competence Center Hagenberg studied three open-source clinical data sets, classifying bone marrow transplant survival, breast cancer malignancy, and heart failure mortality[55]. In addition, similar patients were clustered with a quantum approach by researchers from the National Institute of Engineering, enabling promising results in predicting heart disease[56].

*Treatments and interventions*

New insights from genomics, clinical research, and diagnostics support a joint goal: effectively treating ill patients and proactively keeping people healthy. Ideally and in line with precision medicine, all of this happens at the level of an individual. As for diagnostics, quantum AI/ML approaches are well-suited for the types of calculations encountered here.

Analyzing millions of anonymized electronic health records (EHRs), researchers from Amgen and IBM demonstrated that quantum algorithms can be competitive with classical methods when predicting persistence for rheumatoid arthritis patients, that is, the time interval from start to discontinuation of a given therapy[57]. In another study by an international team of researchers from European Quantum Medical, Orthopaedic Specialists, the George Emil Palade University of Medicine, and the University of Ferrara, treatment effectiveness of knee arthroplasty was assessed based on clinico-demographic data from 170 individuals who had undergone treatment over two years[58]. COVID-19 outbreaks were predicted with a quantum algorithm by a team from the Delhi Technological University and CHRIST through the number of confirmed cases, number of deaths, and number of recovered individuals[59].

> Author case study 1
> Current NISQ systems and quantum (AI/ML) algorithms are generally (not yet) suited to problems that require large volumes of classical data, as noted earlier; overcoming this restriction is an active area of research[60]. Therefore, careful preprocessing is often required to select the right subsets of data for quantum analysis. Moreover, applications where the classical data sets tend to be small naturally lend themselves to quantum computing. For instance, there are clinical use cases that are based on only dozens of samples such as clinical trial and translational medicine cohorts for rare diseases or very specific conditions. The author and collaborators from Amgen and IBM built such an end-to-end preprocessing pipeline for EHRs that allowed them to train quantum (and classical) AI/ML models with 5–20 features and 200–300 training samples[57,61]. While, indubitably, it will be important to scale to big data, achieving quantum advantages for smaller data sets will represent a key milestone. This is because there is usually no reason to suppose that such advantages will not persist for larger data sets.

Finally, precision oncology is one of the most promising sub-fields of precision medicine. This is due to the often-significant differences in the characteristics and behavior of cancer cells from patient to patient. As such, there is a need to tailor treatments more; currently, for example, only a third of individuals respond to drug-based cancer therapies[62]. Quantum computing holds promise for precision oncology[63] and in cell-centric therapeutics[64]; it has already been applied in a related field, namely adaptive radiotherapy. Researchers from the H. Lee Moffitt Cancer Center and Research Institute and the University of Michigan were able to achieve competitive quantum results when applying a combination of quantum and classical techniques to an institutional data set based on 67 stage III non-small cell lung cancer patients[65].

**Making it practical and useful**

Given such promising use cases in medicine, it is important to consider how quantum computing can be implemented in practice to achieve maximal usefulness in the current "quantum decade"[66] and beyond. A range of considerations apply.

First, and arguably most importantly, quantum computing necessitates a novel way of thinking about and tackling problems as well as a new set of skills. As mentioned, the hardware and software are fundamentally different and thus quantum applications must be reimagined from the ground up and optimized for the full stack. It takes time to build these abilities and get into a quantum frame of mind. In addition, the journey towards quantum advantages for more and more use cases is likely a continuous one[67]. The depth of quantum knowledge that is appropriate will vary by organization; however, having zero expertise at a healthcare or life sciences organization will at best result in missed opportunities and at worst in stunted innovation as well as a severe competitive disadvantage.

Second, a tool is only as good as its level of practicality. The vision for quantum computing is that of frictionless integration within hybrid quantum-classical computational pipelines. An individual naturally does not need to be concerned how optimal care and medication comes about – all that matters is the optimality. Thus, most people will not even know that quantum computers are operating in the background; they will just notice that certain things have become possible. Furthermore, as the technology continues to mature the focus for medical practitioners applying quantum computing will gradually shift from the intricacies of the hardware to the details of the quantum circuits and algorithms to the development and usage of quantum-enhanced precision medicine applications. All of this is key to ensuring that also the patient and provider experiences are improved, in addition to achieving better health and lower costs, in line with the quadruple aim[68].

> Author case study 2
> Quantum computing is a young technology and as such there are multifarious physical implementations and platforms with associated trade-offs, for instance, concerning execution speeds, noise characteristics, and qubit interconnections. Therefore, it is not only critical to seamlessly integrate quantum and classical systems but also to choose the optimal quantum platform and computer for a given task. This is the focus of a patent where the author and IBM co-inventors disclosed ways to address that challenge, including the application of machine learning to execution results[69].

Third, the high level of regulation in healthcare and the life sciences means that securing approvals and community acceptance is vital for any new technology. For classical AI/ML, tools have been emerging over the last years that help improve the transparency and explainability of models in medicine. As a result, some medical algorithms and devices based on AI/ML have already received Food and Drug Administration (FDA) approvals[70]. Likewise, related methods for quantum models are being developed and will need further refinement to ensure quantum-based algorithms and products can find widespread market acceptance. In parallel, ethical aspects must be considered, including ways to ensure broad access by medical practitioners and individuals to quantum computing systems and breakthroughs, such as new therapies.

> Author case study 3
> Quantum computing and AI/ML have been rapidly developing over the last years. Increasingly, this is becoming a symbiotic relationship; not only does quantum AI/ML enable better models to be more efficiently developed, but AI/ML is improving quantum computing hardware and software[71]. This speed of development makes it even more urgent to address ethical questions in a timely manner. Together with a co-author from the Center for Ethics at the University of Zurich, the author explored why organizations need to take the ethics surrounding quantum and AI technology adoption seriously[72].

Fourth, just as classical AI/ML and other computational methods must have suitable input data, obtaining actionable insights through quantum computing requires access to health-relevant data, including the quickly growing set of real-world data such as information from claims, disease registries, EHRs, and fitness trackers[73]. Thus, it is critical that healthcare data initiatives, for example, standardization, curation, and interoperability, rapidly progress[74]. Scenarios where quantum computing hardware and software have sufficiently matured for many use cases, but the right data are not available, should certainly be avoided.

Ultimately, quantum computing represents another key milestone in medicine's ever-growing collection of tools. True precision medicine, where the right treatments and interventions at the right time are proactively tailored at the individual level[75], is a lofty goal indeed. Quantum computers are expected to drastically accelerate our quest towards this holy grail over the next years and decades (Figure 4). A range of key biological problems that seem intractable today, or which we have not even been able to formulate, will likely become addressable – and we may well end up taking these new quantum-based solutions for granted, perhaps wondering how it was possible to research and live without them.

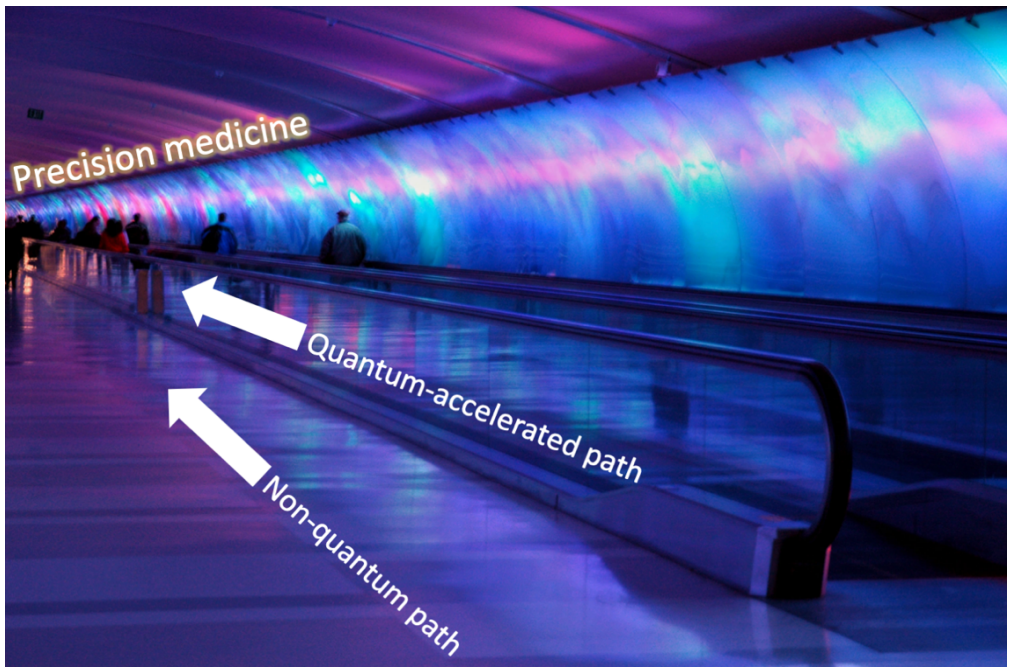

*Figure 4 (adapted from source[76]): Quantum computers are not silver bullets. However, they are poised to dramatically accelerate our progress towards precision medicine.*

The key focus right now is to map quantum algorithms to promising medical use cases, develop proof-of-concept applications, and prepare for novel IT architectures based on quantum-centric supercomputing[77] – all while building and enabling the right skills and getting into that quantum state of mind when thinking about problems. A race is already underway to capture intellectual property and achieve advantages through industry-specific quantum applications, as reflected in the drastic increase in quantum-related papers, patents, and investments over the last years[78]. Succeeding in the quantum age is more akin to a marathon than a sprint, requiring multi-year investments and journeys – and as with any race, the best time to get out of the starting blocks was in the past, and the second-best time is today.